# LOW-COMPLEXITY RECURRENT NEURAL NETWORK-BASED POLAR DECODER WITH WEIGHT QUANTIZATION MECHANISM

*Chieh-Fang Teng* [1], *Chen-Hsi (Derek) Wu* [2], *Andrew Kuan-Shiuan Ho* [2], *and An-Yeu (Andy) Wu* [1]

[1] Graduate Institute of Electrical Engineering, National Taiwan University, Taipei, Taiwan
[2] Department of Electrical Engineering, National Taiwan University, Taipei, Taiwan
jeff@access.ee.ntu.edu.tw, {b03901054, b04901167, andywu}@ntu.edu.tw

## ABSTRACT

Polar codes have drawn much attention and been adopted in 5G New Radio (NR) due to their capacity-achieving performance. Recently, as the emerging deep learning (DL) technique has breakthrough achievements in many fields, neural network decoder was proposed to obtain faster convergence and better performance than belief propagation (BP) decoding. However, neural networks are memory-intensive and hinder the deployment of DL in communication systems. In this work, a low-complexity recurrent neural network (RNN) polar decoder with codebook-based weight quantization is proposed. Our test results show that we can effectively reduce the memory overhead by 98% and alleviate computational complexity with slight performance loss.

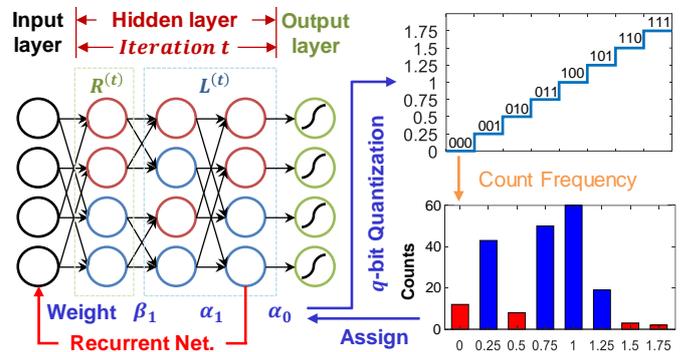

**Fig. 1.** Overview of proposed recurrent neural network polar decoder with codebook-based weight quantization.

## 1. INTRODUCTION

Polar codes, since proposed by Arikan in 2009, have drawn much attention due to their capacity-achieving performance for memoryless channels and low complexity for encoding and decoding [1]. In 2016, 3GPP agreed to adopt polar codes as the official coding scheme for the enhanced mobile broadband (eMBB) control channels of 5G New Radio (NR) [2].

Many polar decoding algorithms have been developed to meet various application requirements. For example, successive cancellation (SC) and belief propagation (BP) are the two most popular decoding algorithms for polar codes. Although SC has a lower complexity, it suffers from high latency as well as limited throughput due to their sequential decoding process [3]-[5]. On the other hand, BP can achieve higher throughput due to its parallelized architecture [5]-[9]. However, the required number of iterative process induces higher decoding complexity. Even with the implementation of min-sum approximation [7], [8] and early termination criterion [7], [9] to reduce complexity and redundant iterations, BP still suffers from unfavorable performance within limited iterations.

To cope with the challenges of BP, deep learning (DL) based decoders have been widely exploited in recent years [10]-[13]. Among the different approaches, they can be mainly divided into two parts. The first one is to completely replace the conventional decoding algorithms with fully-connected neural network [10], [11]. Although [10] shows near optimal performance with structured codes, the training complexity increases exponentially with block length. In addition, the high dimensionality of codewords also requires much more training data and degrades the ability of generalization for unseen codewords [11]. In contrast, [12] and [13] leverage the well-developed BP algorithm via unfolding the iterative structure to the layered structure of neural network. In addition, the weights are also assigned to the connections of the architecture to achieve better performance. By adopting these methods, [12] and [13] can improve the convergence rate and reduce the required number of iterations for Bose-Chaudhuri-Hocquenghem (BCH) codes and polar codes, respectively. However, two issues should be addressed:

1) *Additional memory overhead*: Although the technique of deep learning is very powerful, the massive number of weights results in considerable memory overhead for storage, which hinders the deployment of deep neural networks in communication systems.

2) *Massive multiplications*: Compared with conventional BP algorithm, [12] and [13] have better performance, the weights on the neural networks suffer from massive multiplications and lead to an increase in complexity and occupied area for hardware implementation.

In this paper, by using the design concept of DL, we propose a novel recurrent neural network polar decoder with codebook-based weight quantization. Fig. 1 shows an overview of our proposed architecture. Our main contributions are summarized as follows:

1) Recurrent architecture is adopted to force the neural network to learn the shareable weights among different iterations of BP algorithm, which can effectively reduce memory overhead by 80%.

2) A fixed point codebook for weight quantization is designed, which can further deal with the problem of memory storage by 90%. In addition, it enables us to replace multiplications with additions, which makes the neural network decoder more feasible for hardware implementation.

This research work is financially supported in part by National Taiwan University and MediaTek Inc., Taiwan, under grants NTU-07HZA38001 and MTKC-2018-0167. The first author is also sponsored by MediaTek Ph.D. Fellowship program.

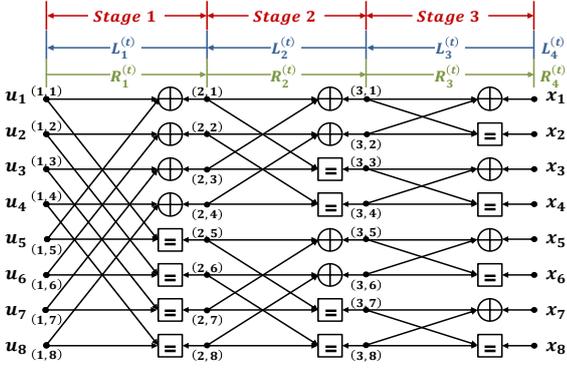

**Fig. 2.** Factor graph of polar codes with $N = 8$.

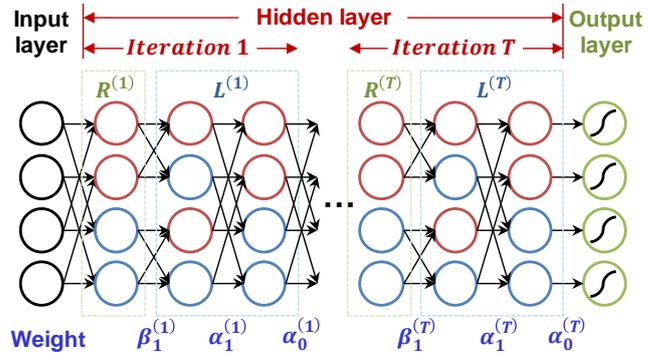

**Fig. 3.** Deep neural network-based belief propagation decoding with $N = 4$.

The rest of this paper is organized as follows. Section **2** briefly reviews the DL-based decoder. Section **3** illustrates the proposed recurrent architecture and the design of codebook for weight quantization. The numerical experiments and analyses are shown in Section **4**. Finally, Section **5** concludes our work.

## 2. PRELIMINARIES

### 2.1. Polar Codes with Belief Propagation Decoding

To construct an $(N, K)$ polar codes, the $N$-bit message $\boldsymbol{u}^N$ is recursively constructed from a $2 \times 2$ polarizing transformation $\boldsymbol{F} = \begin{bmatrix} 1 & 0 \\ 1 & 1 \end{bmatrix}$ by $\log_2 N$ times to exploit the channel polarization [3]. As $N \to \infty$, the synthesized channels tend to two extremes: the noisy channels (unreliable) and noiseless channels (reliable). Therefore, the $K$ information bits are first assigned to the $K$ most reliable bits in $\boldsymbol{u}^N$ and the remaining $(N - K)$ bits are referred to as frozen bits with the assignment of zeros. Then, the $N$-bit transmitted codeword $\boldsymbol{x}^N$ can be generated by multiplying $\boldsymbol{u}^N$ with generator matrix $\boldsymbol{G}_N$ as follows:

$$\boldsymbol{x}^N = \boldsymbol{u}^N \boldsymbol{G}_N = \boldsymbol{u}^N \boldsymbol{F}^{\otimes n} \boldsymbol{B}_N, n = \log_2 N. \quad (1)$$

$\boldsymbol{F}^{\otimes n}$ is the $n$-th Kronecker power of $\boldsymbol{F}$ and $\boldsymbol{B}_N$ represents the bit-reversal permutation matrix.

Belief propagation (BP) is a widely used message passing algorithm for decoding, such as low-density parity-check (LDPC) codes and polar codes. The decoding process of polar codes is to iteratively apply BP algorithm over the corresponding factor graph as shown in Fig. 2. For an $(N, K)$ polar codes, there are $n = \log_2 N$ stages and total $N \times (n + 1)$ nodes on the factor graph. Each node $(i, j)$ represents $j$-th node at the $i$-th stage in the factor graph. It has two types of log likelihood ratios (LLRs), namely left-to-right message $R_{i,j}^{(t)}$ and right-to-left message $L_{i,j}^{(t)}$, where $t$ represents the $t$-th iteration. Before the iterative propagation and updating, the values of LLR are first initialized as:

$$R_{1,j}^{(1)} = \begin{cases} 0, & if\ j \in A \\ +\infty, & if\ j \in A^c \end{cases},\ L_{n+1,j}^{(1)} = \ln \frac{P(y_j|x_j=0)}{P(y_j|x_j=1)}. \quad (2)$$

$A$ and $A^c$ are the set of information bits and the set of frozen bits, respectively.

Then, the iterative decoding procedure with the updating of $R_{i,j}^{(t)}$ and $L_{i,j}^{(t)}$ is given by:

$$\begin{cases} L_{i,j}^{(t)} = g\left(L_{i+1,j}^{(t-1)}, L_{i+1,j+N/2^i}^{(t-1)} + R_{i,j+N/2^i}^{(t)}\right), \\ L_{i,j+N/2^i}^{(t)} = g\left(R_{i,j}^{(t)}, L_{i+1,j}^{(t-1)}\right) + L_{i+1,j+N/2^i}^{(t-1)}, \\ R_{i+1,j}^{(t)} = g\left(R_{i,j}^{(t)}, L_{i+1,j+N/2^i}^{(t-1)} + R_{i,j+N/2^i}^{(t)}\right), \\ R_{i+1,j+N/2^i}^{(t)} = g\left(R_{i,j}^{(t)}, L_{i+1,j}^{(t-1)}\right) + R_{i,j+N/2^i}^{(t)}, \end{cases} \quad (3)$$

where $g(x, y) \approx \text{sign}(x)\text{sign}(y)\min(|x|, |y|)$ is the min-sum approximation introduced to reduce complexity [7], [8]. Then, the final estimation of $\hat{\boldsymbol{u}}^N$ after $T$ iterations is decided by:

$$\hat{u}_j^N = \begin{cases} 0, & if\ L_{1,j}^{(T)} \geq 0, \\ 1, & if\ L_{1,j}^{(T)} < 0. \end{cases} \quad (4)$$

### 2.2. Prior Work: Deep Neural Network Polar Decoder

Though min-sum approximation can effectively reduce the complexity, it also induces performance loss. In [13], a deep neural network-based belief propagation (DNN-BP) with DL was proposed, as shown in Fig. 3. This network can be treated as the unfolding structure of polar factor graph with multiple scaling parameters on the connections. Therefore, (3) can be revised as:

$$\begin{cases} L_{i,j}^{(t)} = \alpha_{i,j}^{(t)} \cdot g\left(L_{i+1,j}^{(t-1)}, L_{i+1,j+N/2^i}^{(t-1)} + R_{i,j+N/2^i}^{(t)}\right), \\ L_{i,j+N/2^i}^{(t)} = \alpha_{i,j+N/2^i}^{(t)} \cdot g\left(R_{i,j}^{(t)}, L_{i+1,j}^{(t-1)}\right) + L_{i+1,j+N/2^i}^{(t-1)}, \\ R_{i+1,j}^{(t)} = \beta_{i+1,j}^{(t)} \cdot g\left(R_{i,j}^{(t)}, L_{i+1,j+N/2^i}^{(t-1)} + R_{i,j+N/2^i}^{(t)}\right), \\ R_{i+1,j+N/2^i}^{(t)} = \beta_{i+1,j+N/2^i}^{(t)} \cdot g\left(R_{i,j}^{(t)}, L_{i+1,j}^{(t-1)}\right) + R_{i,j+N/2^i}^{(t)}, \end{cases} \quad (5)$$

where $\alpha_{i,j}^{(t)}$ and $\beta_{i,j}^{(t)}$ denote the right-to-left and left-to-right scaling parameters of $j$-th node at the $i$-th stage during $t$-th iteration, respectively. Then, by utilizing DL techniques, the best combination of $\alpha_{i,j}^{(t)}$ and $\beta_{i,j}^{(t)}$ can be optimized through backpropagation algorithm. In addition, the proposed DNN-BP is actually the general case of min-sum, which demonstrates the expected performance is no worse than conventional BP.

## 3. PROPOSED RECURRENT NEURAL NETWORK DECODER WITH WEIGHT QUANTIZATION

## 3.1. Recurrent Neural Network Polar Decoder

Although the proposed DNN-BP in [13] can address the problem of approximation loss of min-sum with better performance, the deep architecture also induces another problem of intensive memory, namely, the massive memory is occupied by the storage of weights. In addition, it also consumes a lot of energy when fetching the weights. Both of these issues hinder the deployment of DNN-BP in real world communication systems.

Therefore, we propose a recurrent neural network-based belief propagation (RNN-BP) for polar decoder to share the weights among different iterations. This idea was first proposed by [12] for BCH codes, which is effective in reducing significant parameters. However, we further alleviate the memory overhead by combining this idea with the proposed codebook-based weight quantization, which will be introduced in section **3.2**. From the analysis in Sec **4**, we demonstrate a great improvement in memory overhead by combining these two methods.

The architecture of RNN-BP is shown in Fig. 1. We adopt the same factor graph as in [13] for the decoder. Compared with the DNN-BP in Fig. 3, the greatest difference of RNN-BP is that we transform the feed-forward architecture into a recurrent network, which forces the decoder to reuse weights among different iterations and results in a totally different optimization problem. Therefore, we can effectively reduce massive parameters without visible performance degradation. That is, RNN-BP can be seen as a constrained version of DNN-BP with the constrained equation as below:

$$\begin{cases} L_{i,j}^{(t)} = \alpha_{i,j} \cdot g\left(L_{i+1,j}^{(t-1)}, L_{i+1,j+N/2^i}^{(t-1)} + R_{i,j+N/2^i}^{(t)}\right), \\ L_{i,j+N/2^i}^{(t)} = \alpha_{i,j+N/2^i} \cdot g\left(R_{i,j}^{(t)}, L_{i+1,j}^{(t-1)}\right) + L_{i+1,j+N/2^i}^{(t-1)}, \\ R_{i+1,j}^{(t)} = \beta_{i+1,j} \cdot g\left(R_{i,j}^{(t)}, L_{i+1,j+N/2^i}^{(t-1)} + R_{i,j+N/2^i}^{(t)}\right), \\ R_{i+1,j+N/2^i}^{(t)} = \beta_{i+1,j+N/2^i} \cdot g\left(R_{i,j}^{(t)}, L_{i+1,j}^{(t-1)}\right) + R_{i,j+N/2^i}^{(t)}, \end{cases} \quad (6)$$

where $\alpha_{i,j}$ and $\beta_{i,j}$ are independent with $t$ and shareable among different iterations. Finally, the output will be rescaled into the range [0,1] with sigmoid function as below:

$$o_j = \sigma\left(L_{1,j}^{(T)}\right) = \left(1 + e^{-L_{1,j}^{(T)}}\right)^{-1}, \quad (7)$$

and the loss function is cross entropy between transmitted codeword $\boldsymbol{u}$ and output $\boldsymbol{o}$ of RNN-BP:

$$\mathcal{L}(\boldsymbol{u},\boldsymbol{o}) = -\frac{1}{N}\sum_{j=i}^{N} u_j \log(o_j) + (1-u_j)\log(1-o_j). \quad (8)$$

## 3.2. Codebook-based Weight Quantization

Though recurrent architecture effectively reduces the required number of parameters, the floating-point parameters still hinder the hardware implementation of neural network decoder. Besides, the additional multiplications for scaling parameters also increase the computational complexity. Hence, we propose codebook-based weight quantization to effectively quantize the parameters and alleviate computational complexity without visible performance degradation.

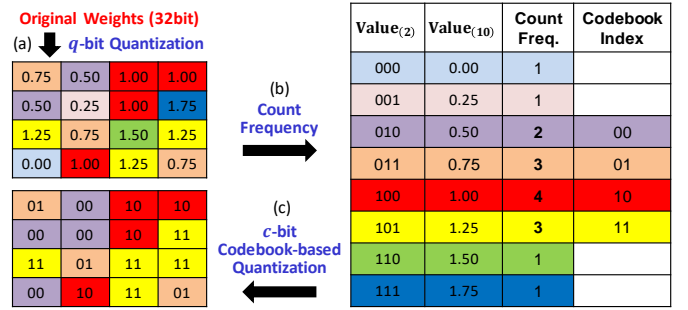

**Fig. 4.** The proposed codebook-based weight quantization: (a) $q$-bit weight quantization from original 32-bit weights with $q = 3$; (b) count the frequency of different quantized weights; (c) further quantize the weights based on the designed codebook with $c = 2$.

In order to maintain the high performance, the weights of our proposed RNN-BP are quantized after each epoch during the training process. As a result, we allow the quantized weights to induce the quantization loss and let the network to learn the best quantized weights. The proposed approach can be mainly divided into two parts, with double quantization to reduce both the required number of weights and the precision for each weights.

The quantization mechanism is illustrated in Fig. 4. Firstly, we quantize the 32-bit floating point weights into $q$-bit fixed point as shown in Fig. 4(a) which is similar to [14]. The weights are rounded to the nearest value that can be represented by $q$-bit. Considering the distribution of weights is positive and close to 1, we set the first bit as integer and other $(q - 1)$ bits as decimal. Thus, the quantization step is $2^{q-1}$. Secondly, we design the codebook based on the quantized weights for weight sharing, as proposed in [15], to further reduce the number of distinct weights among the network.

In contrast to [15], since the weight variation is already limited by $q$-bit quantization, there are only $2^q$ types of weights. Therefore, the selection of codebook weights is simply done by counting the frequency of the quantized weights. When the size of the codebook is $c$ bits, or $2^c$ distinct values, we can choose the $2^c$ most common values to enter into the codebook as shown in Fig. 4(b). As the weights are quantized to the nearest value in the codebook, the memory used for each weight is further reduced from $q$ bits to $c$ bits with the simple binary index assignment as shown in Fig. 4(c).

## 4. SIMULATION RESULTS

The simulation setup is summarized in Table I.

TABLE I. SIMULATION PARAMETERS

| Encoding | Polar code (64,32) |
|---|---|
| Signal to Noise Ratio (SNR) | 0, 1, 2, 3, 4, 5 |
| Training codeword/SNR | 40000 |
| Testing codeword/SNR | 100800 |
| Mini-batch size | 2400 |
| Optimizer | RMSProp |
| Training and testing environment | DL library of TensorFlow with NVIDIA GTX 1080 Ti GPU |

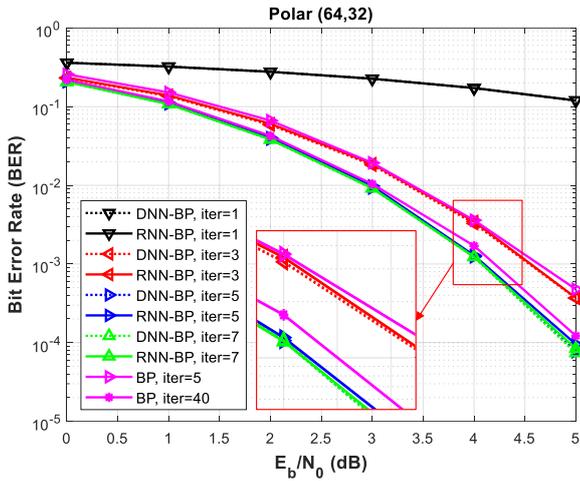

**Fig. 5.** Comparison of BER performance between the proposed RNN-BP and prior works under different BP iterations.

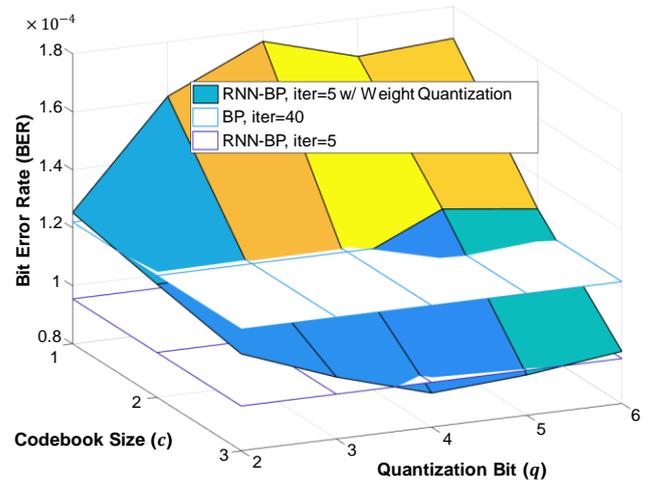

**Fig. 6.** BER performance of proposed RNN-BP with codebook-based weight quantization under different settings.

### 4.1. Performance of NN-based Polar Decoder

The first experiment is to simulate the bit error rate (BER) performance under different BP iterations, which has great impact on the computational complexity and the required number of parameters for neural network decoder. We compare the proposed RNN-BP with conventional BP [7] and DNN-BP [13]. The simulation results are shown in Fig. 5. Firstly, we can observe that both DNN-BP and RNN-BP converge and achieve a great performance with just 5 iterations. On the other hand, conventional BP demands more than 40 iterations to achieve the same level as the neural network decoder. Therefore, conventional BP consumes more computational resources with higher energy consumption and longer latency. Moreover, RNN-BP has almost the same BER as DNN-BP with 80% reduction in parameter count, which makes RNN-BP more feasible for deployment in communication systems.

### 4.2. RNN-BP Decoder with Codebook-based Weight Quantization

To further reduce the memory storage and computational complexity, we compare the performance under different criterions of weight quantization to evaluate the tradeoff between performance and hardware consumption and determine the most optimized quantization parameters. The simulation results are shown in Fig. 6. The codebook size $c$ is range from 1 to 3, and the quantization bit $q$ is ranged from 2 to 6. To our surprise, when codebook size is $c = 1$, the BER deteriorates as the quantization bit increases. We speculate that this is because weight quantization may change the error surface of optimization problem and result in a convergence to local minimum earlier. As the codebook size $c$ greater than 2, the longer quantization bits, the better BER. In addition, we observe that the fixed point RNN-BP even outperforms its floating point counterpart, which may be caused by weight quantization acting as a regularization measure, thus contributing positively to the performance. Therefore, we can effectively reduce memory overhead without visible performance degradation.

TABLE II. ANALYSIS OF COMPUTATIONAL COMPLEXITY

| | **Addition** | **Multiplication** | **Memory (bit)** |
|---|---|---|---|
| BP [7] | $2TN\log N$ ~30,720 (100%) | 0 | 0 |
| DNN-BP [13] | $2TN\log N$ ~3,840 (12.5%) | $2TN\log N$ ~3,840 (100%) | $64TN\log N$ ~122,880 (100%) |
| Proposed RNN-BP with codebook-based weight quantization | $2qTN\log N$ ~15,360 (50%) | 0 | $2cN\log N$ ~2,304 (1.875%) |

Further analysis of hardware complexity is shown in Table II. We compare the proposed approach with conventional BP [7] and DNN-BP [13]. The required numbers of iterations $T$ for BP, DNN-BP, and RNN-BP are based on previous simulation results and set to 40, 5, and 5, respectively. Firstly, we can observe that DNN-BP dramatically reduces the addition operations at the expense of multiplication operations and significant memory overhead. However, with recurrent architecture and codebook-based weight quantization, we effectively reduce the memory overhead by 98% with codebook size $c = 3$. In addition, the quantized weights can help to reduce the operation of multiplications 100% by replacing multiplications with $(q - 1)$ times additions and $q = 4$, which occupies less area for hardware implementation and maintains improvement over the original BP method.

### 5. CONCLUSION

In this paper, we present a novel recurrent neural network polar decoder with codebook-based weight quantization. It can learn the shareable parameters with effective reduction of memory overhead. Moreover, the proposed weight quantization can further alleviate hardware complexity without performance degradation. Our proposed design is low complexity and feasible for realizing neural network decoders in communication systems.